\documentclass[english, 12 pt]{article}
\usepackage[T1]{fontenc}
\usepackage[latin9]{inputenc}
\usepackage{graphicx}
\usepackage{babel}
\usepackage{multicol}
\usepackage{amsmath}
\usepackage{rotating}           
\usepackage{color}

\makeatletter

\newcommand{\fig}[1]{Figure (\ref{#1})}

\newcommand{\eq}[1]{Eq.(\ref{#1})}

\newcommand{\se}[1]{Sec. (\ref{#1})}

\newcommand{\la}[1]{ \label{#1}}
\renewcommand{\a}{\alpha}

\newcommand{\e}{\epsilon}

\renewcommand{\r}{\mathbf{r}}

\newcommand{\bsubs}{\begin{subequations}}
\newcommand{\esubs}{\end{subequations}}

\providecommand{\ba}{\begin{align}}
\providecommand{\ea}{\end{align}}

\usepackage{multirow}
\newcommand{\be}{\begin{equation}}
\newcommand{\ee}{\end{equation}}

\newcommand{\bea}{\begin{eqnarray}}
\newcommand{\eea}{\end{eqnarray}}

\begin{document}

\title{ Slippery Wave Functions V2.01  \\
 }

 \author{ Leo P. Kadanoff\footnote{e-mail:  lkadanoff@gmail.com}~
 \\
\\
 The James Franck Institute\\
The University of Chicago
\\ Chicago, IL USA
\\ 
and \\
The Perimeter Institute,\\
Waterloo, Ontario, Canada \\
\\}

\maketitle
\begin{abstract}
Superfluids and superconductors are ordinary matter that show a very surprising behavior at low temperatures.   As their temperature is reduced, materials of both kinds can abruptly fall into a state in which they will support a persistent, essentially immortal, flow of particles. Unlike anything in classical physics, these flows engender neither friction nor resistance. A major accomplishment of Twentieth Century physics was the development of an understanding of this very surprising behavior via the construction of partially microscopic and partially macroscopic quantum theories of superfluid helium and superconducting metals.   Such theories come in two parts:  a theory of the motion of particle-like excitations, called quasiparticles, and of the persistent flows itself via a huge coherent excitation, called a condensate.  Two people, above all others, were responsible for the construction of the quasiparticle side of the theories of these very special low-temperature behaviors: Lev Landau and John Bardeen.  Curiously enough they both partially ignored and partially downplayed the importance of the condensate.  In both cases, this  neglect of the actual superfluid or superconducting flow interfered with their ability to understand the implications of the theory they had created. They then had difficulty assessing the important advances that occurred immediately after their own great work.  

Some speculations are offered about the source of this unevenness in the judgments of these two leading scientists.   
    
\end{abstract}

\newpage

\tableofcontents

\newpage{ }
\section{Introduction}
In physics publications, one  relatively rarely sees direct statements saying that contemporary authors are wrong.   So it is particularly striking when the authors of two of the most important papers in condensed matter physics go out of their way to criticize other physics papers, saying flatly that they are far from the mark.  We should look even more carefully when the two criticisms appear to be quite similar, and also somewhat problematic from the perspective of today's understanding of physics.\footnote{A good general reference to the development of this part of science is\cite{Lillian}.  A predecessor that carefully assesses the helium work is S\'{e}bastien Balibar's {\em The Discovery of Superfluidity}\cite{Balibar}. When he and I overlap, our conclusions are much the same and were reached independently.   The outline for much of this paper was suggested by the late Allan Griffin\cite{AG} in his  introduction to a Varenna volume on Bose-Einstein condensation. Mistakes and errors of judgment are my own rather than Allan's, of course.}

The two authors are Lev Landau in the 1941 paper entitled\footnote{{\em Helium II}\/ is the name for the state of helium below the temperature for its transition to superfluidity.} {\em The Theory of Superfluidity in Helium II}\/\cite{L1}   and the collective authorship of John Bardeen, Leon Cooper, and Robert Schrieffer, abbreviated BCS, in their blockbuster papers on the theory of superconductivity\cite{BCS1,BCS2}.

\subsection{Background}\la{Background}
Landau's paper\cite{L1} was inspired by the 1938 discovery of superfluid motion in the natural form of helium\cite{Kapitsa,AM} when that material is taken to sufficiently low temperature.  In this surprising motion, helium can move through small cracks and in thin sheets without friction.  If put into a circular channel( See \fig{RaceTrack}), it can flow around apparently forever without slowing down.

Superconductivity, discovered by Heike Kamerlingh Onnes\cite{HKO}  in 1911, exhibits frictionless flow, like that of superfluid helium, but in this case occurring in metals like mercury and aluminum.   However, in superconductors electrons are in motion. These charged particles engender an electromagnetic field that tends to push\cite{MO} magnetic fields out of the superconducting body. 

As the temperature is lowered both superfluidity and superconductivity appear quite abruptly at low temperatures in a change of material behavior called a phase transition\cite{LPK09,LPK11,LPK10}.

By 1941 some progress had been reached in understanding superconductivity. 
A good phenomenological theory had been developed by Gorter and Casimir\cite{GC} and had been extended to include electromagnetic effects by Fritz and Heinz London\cite{FHLondon}.  The Gorter-Casimir work included a ``two-fluid'' analysis that considered the superconductors to be composed of two interpenetrating fluids: a normal fluid and a superfluid.  The latter was given all the anomalous properties, including the ability to flow without friction.  These fluids were not in any obvious sense quantum.   Fritz London, on the other hand, suggested that the superconductor was behaving like a huge atom in which the existence of quantization forced a special rigidity upon the wave function for the low temperature state\cite{1935d}.   London argued that this rigidity tended to reduce the magnetic field within the atom in the same way as the current flow in the superconductor reduced its internal magnetic field\footnote{The technical name for the magnetic response with this kind of rigidity is {\em diamagnetism.} So London is saying that a superconductor is like a big diamagnetic atom.}. 

By 1941, some work had also been done on the superfluid behavior of helium.    There was an obvious analogy between the eternal currents produced by superconductivity and superfluidity.
 Tisza had suggested that a helium could also be described by a two-fluid model\cite{Tisza}, rather like the one that applied to superconductors.   The work of Fritz London and Tisza suggested that perhaps superfluid helium could be likened to the anomalous state of a non-interacting boson system, as had previously been described by Albert Einstein\cite{Einstein}.\footnote{The interaction between Tisza and Fritz London in the period 1937-1939 is described in detail in \cite{Balibar}[pages 454-456].}

 In 1924, Einstein\cite{Einstein} had developed a theory of the behavior of a non-interacting gas of particles obeying the kind of quantum symmetry first posited by Satyendra Nath Bose\cite{Bose}.   In Einstein's theory, the gas is composed of two parts:
\begin{itemize}
\item  A spectrum of independently moving particles, with energy-momentum relation $\e=p^2/(2m)$, as is appropriate for   noninteracting, nonrelativistic particles.
\item  A condensate, a single-particle quantum state with wave function $\Psi(\r)$, occupied by a finite fraction of all the particles in the system\footnote{In this paper, I discuss the condensate as if it were always one and only one mode of oscillation, macroscopically occupied. This picture applies in three or higher dimensions. In two dimensions, however, the condensate is spread out over a whole collection of modes.}  
\end{itemize}
 Einstein's results imply  that his phase transition occurs because a finite fraction of all the particles in the fluid fall into a single quantum state, which then might be described by a single wave function $\Psi(\r).$ The Bose (or Bose-Einstein) character of the particles permit many different particles to be described by a single wave function.  That would be impossible for the other kind of quantum particles, ones that obey Fermi-Dirac system.

\begin{figure}
\begin{multicols}{2}
\includegraphics[height=6.5 cm ]{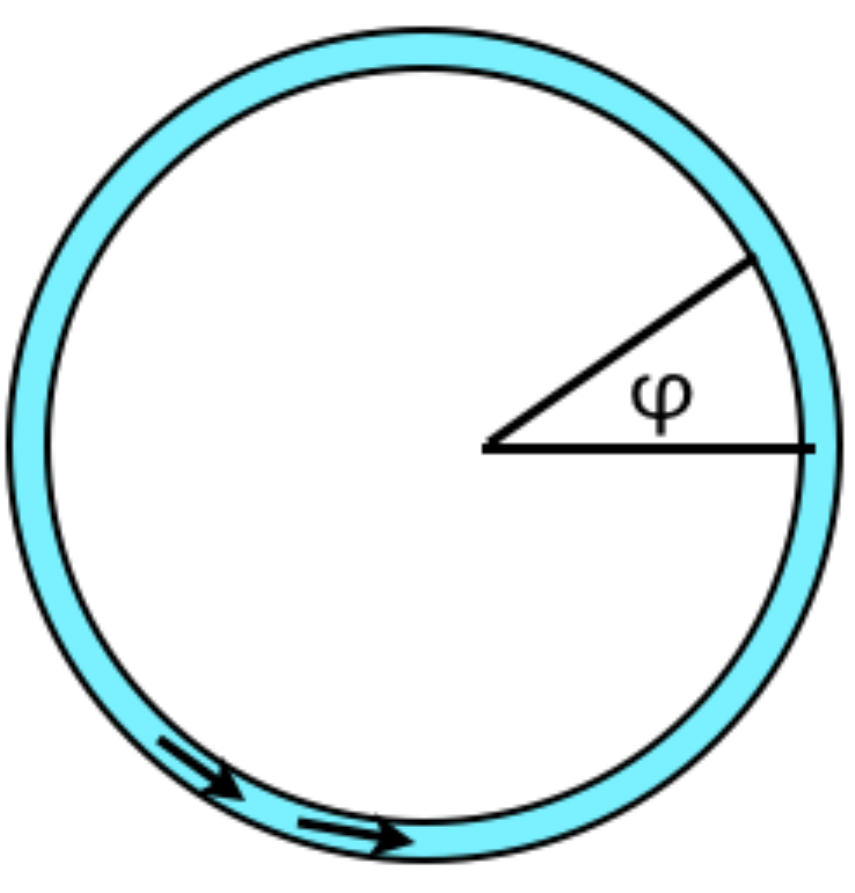}
\caption{Superflow.  Helium is contained in a narrow circular channel.  It is set in circular motion at a temperature below its $T_c$ for the transition to superfluidity. The flow will continue essentially indefinitely.  In describing its quantum behavior we use a angular coordinate, $\phi$.  }
\la{RaceTrack}
\end{multicols}
\end{figure} 

However, a very high  barrier separated physicists from an acceptable theory of these superflows.  Felix Bloch had proven that the lowest energy state of {\em any}\/ system, including multiply connected material like that in \fig{RaceTrack} had to have current equal to zero.  But superflows could be seen to persist down to very low temperatures so that we might expect their low-temperature configurations to be very much like those in the ground state.  There seemed to be no way out of this dilemma.  
\section{Landau: 1941, 1948}
\subsection{Quasiparticle analysis}Landau's 1941 paper\cite{L1} grapples with many ideas that would subsequently form the center of condensed matter physics.  For example, the paper extends concepts of elementary excitations to include excitations in low temperatures helium. Previous work on elementary excitations in normal metals, done by Felix Bloch\cite{Bloch} and others\cite{Hoch},   had the excitations given properties closely based upon the properties of a non-interacting gas of particles. As in the gas, the excitations would each have well-defined  momentum, ${ \bf p}$, and the excitation energy was then taken to be a function of the momentum, $\e({\bf p})$.  Excitations like these would later\cite{Bogoliubov} be called  ``quasiparticles.''\footnote{An {\em quasiparticle}  is a long-lived, particle-like, excitation in a many-body system. Some people would reserve the word for a fermionic mode and call a bosonic  mode a collective excitation.  Since the theory of the two kinds of excitations are quite similar,  I here use quasiparticle to refer to both.}    Their behavior was to be analyzed in terms of the Hamiltonian mechanics of their independent motion. Also included were the occasional collisions of the quasiparticles.   This point of view became quite celebrated through Landau's later application\cite{L3} of quasiparticle concepts to produce a theory of the low temperature behavior of the fluid composed of the helium isotope with atomic weight three,  $^3$He.  

It is important to note that Kapitsa's superfluid and Onnes' superconductors are formed from very different constituents. The material studied by Kapitza\cite{Kapitsa} and   Landau-1941\cite{L1,L2} is natural helium,  mostly composed of the isotope $^4$He, obeying Bose-Einstein statistics. On the other hand the electrons in metals and  the isotope $^3$He, later studied by Landau\cite{L3},  obey Fermi-Dirac statistics.  In the Bose case, the wave function for the system will be symmetrical under an interchange of the coordinates of any two particles. However, for Fermi particles the wave function for the system will change sign under such an interchange.  This difference between the two kinds of particles is likely to manifest itself quite strongly at low temperatures,  where quantum properties will be important.  As mentioned by Landau, the statistics will determine the detailed scattering and correlation properties of the quasiparticles.  But since Landau does not focus upon a condensate, he cannot make use of the fact that condensates are directly possible for Bose particles but can  only occur for Fermi particles if they first group together to form ``molecules'' with even numbers of fermions.    

A major portion of the 1941 Landau paper\cite{L1} is  devoted to a discussion of what we would now call the quasiparticle spectrum of superfluid helium and how that spectrum is connected with the observed superfluidity.   An important element of Landau's analysis is his suggestion of a new kind of quasiparticle excitation, given the name, ``roton'', which has the important property that its energy, $\e(p)$, has a non-zero minimum value, $\Delta$, at a particular value of $p$.   

Most important, Landau here gives three different arguments for superfluidity.  The first of these is the famous {\em Landau criterion}.  A material moving through a tube can lose momentum to the walls containing it if it is moving fast enough.    According to this paper, momentum can be lost if the velocity is greater than 
\be
v* = \text{minimum~} [ \e({\bf p})/p]
\la{LandauCriterion}
\ee
where the minimum is taken with respect to ${\bf p}$.
If the minimum is zero, as it is for example in the case of non-interacting particles, then there will be no superfluidity.   

Since Landau does not really tell us why superfluidity exists, it is hard to tell from his arguments the extent to which his criterion is true.\footnote{Recently G. Baym and C. Pethick\cite{BaP} have argued that the Landau criterion is neither necessary nor sufficient for superfluidity by pointing to counterexamples of both types.  They assert that the criterion does work to describe the possible reduction of superfluidity via the loss of momentum in collisions, but that this loss can have the modest result of converting a fraction of the superfluid component of the liquid into the normal fluid rather than the dramatic result of destroying the superfluidity.} However, given the many mechanisms for broadening the distributions of both energy and momentum, it seems very implausible that any condition like that implied by \eq{LandauCriterion} can begin to account for the very long-lived nature of the flow of superfluid helium.

Secondly, Landau points out that if the superfluid flow velocity can be constructed as the gradient of some scalar quantity, then the superflow cannot produce a net force on any solid body.  However, the quasiparticle microscopics discussed by him does not translate into a gradient superflow.  

\subsection{Superconductivity}
The last requirement for superfluidity is given in Landau's final section, titled ``The Problem of Superconductivity.'' This section starts out by saying that superconductivity and superfluidity are closely related phenomena. It is thus not quite clear why Landau has shifted his focus from helium to metals.  Landau mentions the previous suggestion, ``advanced several times''  that superconductivity arises from a energy gap in its spectrum.  This paper previously argued that just this kind of gap, in the roton spectrum, was the basic cause of superfluidity in helium.  But then he avers that the criterion of \eq{LandauCriterion} is a necessary but not sufficient condition for superflow.  One needs a special behavior of the entire system, a behavior that Landau says, incorrectly, is quite different in superconductor and superfluid.  That special behavior includes motions\cite[page 202]{L1}  ``in which the liquid moves as a whole, in a macroscopic manner.''  Landau only defines that special behavior at zero temperature, in which he says that for weak magnetic fields  the ground state wave function should be multiplied by a factor of the form   
\be
\prod_\a ~ \exp [ i \chi(\r_a)/\hbar] .
\la{factor}
\ee
where $\chi(\r)$ is proportional to the vector potential, $\bf{A(\r)}$, describing the applied magnetic field at the spatial point $\r$. 
See \cite[equation 9.1]{L1}, expressed in slightly different form.  This is exactly what we would say today.   But it is also the rigidity statement of Fritz London\cite{1935d}.  For some reason, Landau fails to mention that this step is old hat.   He does not, in fact, refer to London at all.    

Landau avoids the next step which we, today,  might want him to take: to define a wave function describing the common flow properties of a superconductor in a magnetic field as 
\be
\Psi(\r) =  \exp [ i \chi(\r)/\hbar], 
\la{condensate}
\ee
where $\chi(\r)$ might be complex,     
We might then wish Landau to say that the existence of such a wave function is the essential requirement for superfluidity.\footnote{To complete the argument by describing the magnetic perturbation that might exist at non-zero temperatures, one would envision multiplying both sides of a quantum density matrix by factors  like that in \eq{factor}. This step is taken in a later paper by Ginzburg and Landau\cite{GL}  described in \se{Ginzburg-Landau} below.}

The issue in this note is not only these criteria for superfluidity, but also why Landau  does not touch upon using a wave function in his  requirements for superfluidity.   Some insight is perhaps derivable from asides in which Landau goes out of his way to argue that a paper by Laszlo Tisza\cite{Tisza} is incorrect.  Tisza is responsible for introducing a two-fluid model of superfluid helium\cite{Tisza0} as distinct from the two-fluid model of superconductivity.  In a later paper\cite{Tisza}, he makes the insightful but not-very-audacious claim that the observed superfluidity of $^4He$, a boson fluid,  is related to the Bose-Einstein theory for condensation of non-interacting bosons. 
Tisza further identifies the perfect superfluid with the condensed part of the Bose-Einstein construction.  A natural generalization of the theories previously used for electronic excitations\cite{Bloch,Hoch}  would be to use the non-interacting Bose gas as a basis for constructing a theory of superfluid helium.  In that case, both a quasiparticle spectrum and a condensate wave function would appear in the helium theory.\footnote{This strategy of taking the noninteracting system to be the template for the construction of a quasiparticle theory, the latter being modified to include a more general energy momentum relation, is exactly the one later followed by Landau in his later construction of a quasiparticle theory of $^3$He\cite{L3}.     }   Indeed, the condensate wave function is regarded  today as a crucial part of the superfluid behavior of helium.    But, back to Landau, who in 1941 took a different view.

\subsection{Landau and Tisza}
Landau mentioned Tisza's ideas twice in this paper.  On the very first page of his article, he notes that Tisza argued that superfluid helium should be considered to be a Bose gas with a finite fraction of their particles in a state of lowest energy. Further,   
 \begin{quotation}
...[Tisza] suggested that the atoms found in the normal state (a state of zero energy) move through the fluid without friction. This point of view, however, cannot be considered as satisfactory. Apart from the fact that liquid helium has nothing to do with an ideal gas, atoms in the normal state would not behave as ``superfluid''. On the contrary, nothing would prevent atoms in the normal state from colliding with excited atoms, i.e. when moving through the liquid  they would experience a friction and there would be no superfluidity at all. In this way the explanation advanced by Tisza has no foundation in his suggestions but is in direct contradiction to them.\footnote{Part of Landau's trouble with Tisza seems to be contained in the word ``normal'' as in ``normal fluid''.  They appear to use the word differently so that what Tisza is saying is hard for Landau to interpret.}
\end{quotation}
After he discusses the roton excitation and its energy-minimum at energy equal to $\Delta$,  Landau presents a footnote\cite[page 192]{L1} 
\begin{quotation}
It must be mentioned that for an ideal gas $\Delta=0$ and, therefore, even at absolute zero it would not disclose the phenomenon of superfluidity at any velocities of the flow contrary to Tisza's suggestion.
\end{quotation} 
These words suggest an absolute rejection of Tisza's thesis that what we now call the condensate, having a huge number of  particles described by the lowest energy wave function, has anything to do with superfluidity.   Landau's position is somewhat delicate in that he certainly accepts the two-fluid model of Tisza and H. London\cite{HLondon} as a descriptor of superfluidity.   But the two-fluid model is phenomenological and  macroscopic in content.  Landau would like to reject out of hand the connection of Einstein's  microscopic theory with any description of superfluid helium.    In the Twenty-First Century, we might consider this point of view to be quite remarkable since we now consider the superfluid motion to be describable by this condensate wave function. Further, we might remark that this identification of superfluid motion is in reasonable measure based upon the work of Landau himself, specifically but not exclusively in his Ginzburg-Landau\cite{GL} paper that will form a later focus of this paper. 

Landau returned to Helium II at least twice more.  In 1947\cite{Landau1947} he wrote a two-page paper in which he used experimental data to correct an error he had made in placing the roton minimum at zero momentum. As a result Landau's revised picture of the quasiparticle spectrum became quite similar to the Feynman picture\cite{Feynman} mostly held today.  

In addition, Landau has a 1948 short paper\cite{L2} that is mostly devoted to assessing the progress made since 1941 in understanding superfluidity of helium.  It does not have the magnificent new ideas of the 1941 work.     Landau does point out the recent experimental and theoretical work that has supported his previous predictions while he continues to disagree with Tisza about the form and substance of the latter's work.  Once more Landau suggests that the quasiparticle spectrum determines the entire behavior of the helium. He hardly mentions the condensate.    Here, Landau devotes six footnotes to Tisza.  The last sentence of the first  is
\begin{quotation}
However, his entire quantitative theory (microscopic as well and thermodynamic-hydrodynamic) is in my opinion, entirely incorrect.  
\end{quotation}
The paper continues in the same vein in the text.
\begin{quotation}
The hydrodynamic equations given by Tisza are, in my opinion, quite unsatisfactory. It is easy to see that in their exact form they even violate the conservation laws!.  
\end{quotation}
 
Landau also mentions N.N. Bogoliubov's important work\cite{Bogoliubov} on weakly coupled Bose gases.  Landau mentions that Bogoliubov has determined the ``general form'' of the quasiparticle theory, while saying that Bogoliubov's work ``does not have any direct bearing on the actual liquid Helium II.''  Further Landau repeats his old assertion that the Landau criterion is sufficient for the deduction of superfluid flow behavior, while ignoring Bogoliubov's deduction of the existence of a condensate.   It is, of course, the last point that will prove crucial for understanding superflow.   Landau's rhetoric about both Tisza and Bogoliubov has quite a strong tone, perhaps indicative of some extra-scientific reasons for his argument.

I further discuss Landau's motivation  in \se{NotInverntedHere} below.

\section{BCS: 1957}    
One and a half decades  later, the blockbuster BCS work was published, first as a letter\cite{BCS1} and then in full form\cite{BCS2}.    As we shall see,  these BCS publications were similar to Landau's work in showing an unwillingness to admit the possibility that a condensate wave function might play an essential role in superflow.  

These BCS papers do contain the first satisfactory quasiparticle theory of superconductivity. Over the decades between the 1911 discovery of superconductivity by Heike Kamerlingh Onnes\cite{HKO} and the BCS work, theorists had put together a phenomenological theory of superconductivity, but there had never been anything close to a successful microscopic theory.   The phenomenological theory\cite{FHLondon} included equations describing the current flow that described the most important macroscopic facts about the electromagnetism of superconductivity: \begin{itemize}
\item Electrical currents could flow without ohmic resistance.
\item They could flow apparently forever around a ring. 
\item These flow properties appeared abruptly at a critical temperature, $T_c$ and persisted below that temperature.
\item Current flow in a pure superconductor eliminated the magnetic field from within the material.   This is called the Meissner\cite{MO} effect.
\end{itemize}

In addition to this phenomenological work, there were several attempts at constructing a microscopic theory.  None of these described in detail the behavior of real superconducting materials.   For this note, the most important previous work was a series of studies by M. R. Schafroth, S, T, Butler and J. M. Blatt, particularly \cite{SBB},  which pointed out a possible analogy between superfluid helium and superconductors, including the existence of a condensate which might drive the flow.    

\subsection{From Cooper pairs to quasiparticles}
The BCS papers were based upon an earlier work in which Leon Cooper\cite{Cooper} worked with an attractive interaction between electrons produced by phonons that had previously been derived by Bardeen and David Pines\cite{BardeenPines}.   Using this interaction, Cooper showed that the interaction produced a tendency   for pairs of electrons to form bound pairs. These pairs had opposite spins directions and momenta  roughly equal in magnitude and and opposite in direction.  

The existence of a possible boson interpretation of superconductivity was evident in the Cooper letter which stated 
\begin{quotation}
It has been suggested that superconducting properties would result if electrons could combine in even groups so that the resulting aggregates would obey Bose statistics$^{a,b}$.
\begin{itemize}
\item a. V.L. Ginzburg {\em Uspekhi Fiz. Nauk. }{\bf 48}, 25b (1957), 
\item b. M.R. Schafroth {\em Phys.  Rev.} {\bf 100}, 463 (1955) 
\end{itemize}
\end{quotation}
 Indeed it is well known that pairs of identical fermions, viewed from afar, would appear to obey Bose-Einstein statistics.
The argument is much older than 1955, and goes back at least to Richard Ogg's 1946 pointer\cite{Ogg} to Bose-Einstein condensation as a source of superconductivity.

Building upon Cooper's work,  BCS  showed that when there was a multiplicity of Cooper pairs in a metallic situation, the electrons in the metal fell into a state substantially different from that of the usual band theory\cite{Bloch,Hoch,BP}.  In that theory, electrons in a metal are taken to behave rather similarly to a non-interacting gas of particles that obey Fermi-Dirac statistics. They form quasiparticles labeled by particle momentum and spin.

In the BCS paper,  the excitations are once more quasiparticles labeled by spin and momentum.  The major difference is that, as in the Bogoliubov quasiparticles for Helium II\cite{Bogoliubov},  the quasiparticle is a linear combination of a particle with momentum ${\bf p}$ and a hole\footnote{BCS could have reminded the reader that the 1946 paper of N. N. Bogoliubov\cite{Bogoliubov} contained a very similar form of excitation. They did not do so.  This failure pushed the reader away from looking at the analogy between superfluids and superconductors } with momentum ${-\bf p}$.
In addition the energy spectrum of the excitations is strongly modified from the structure of the band theory.  The BCS quasiparticles have an energy gap,  given the value $\Delta$.  The paper then briefly argued that this modified gas of excitations  could display the frictionless flow of current that was the first described property of superconductors.  In part, the argument was that the gapped excitation spectrum satisfied a form of the Landau criterion, appropriately modified for Fermi-Dirac particles.

\subsection{Looking away from Bose-Einstein behavior}

One can see the viewpoint of BCS in the introduction to the main paper.  They start by listing the facts that should be explained by a theory of superconductivity.  The first fact is that superconductivity is produced by a second order phase transition\footnote{It is important to the BCS arguments that the transition not be first order in nature.  First order phase transitions permit and entail jumps in behavior.  Second order ones introduce new behavior at the phase transition, but do so gradually. BCS expect continuity at the phase transition. }.   They continue with the ``evidence for an energy gap'', and the expulsion of the magnetic field from the interior of the bulk superconductor.  Only in fourth place is ``effects associated with infinite conductivity''.  Last comes the dependence of $T_c$ on the isotopic mass of the ions.   Thus, the actual infinite conductivity is not by any means of first importance to this work.    

The BCS work shies away from discussing a condensate wave function and the previously suggested connection of superconductivity with bose behavior\cite{Ogg, Sha, Sha1,SBB}. 
The first, and only, reference to the behavior of a Bose-Einstein gas is in footnote 18, elicited by a mention of [quantum] coherence of a large number of electrons.  The footnote reads
\begin{quotation}  
Our picture differs from that of Schafroth,  Butler, and Blatt, {\em Helv. Phys. Acta} {\bf 30}, 93 (1957) who suggest that pseudomolecules of pairs of electrons of opposite spin are formed.  They show that if the size of the pseudomolecules is less than the average distance between them, and if other conditions are fulfilled, the system has properties similar to a charged Bose-Einstein gas, including a Meissner effect and a critical temperature of condensation.  Our pairs are not localized in that sense, and our transition is not analogous to a Bose-Einstein transition. 
\end{quotation}

This statement points the reader away from any analogy with  Bose-Einstein condensation.  However, to be fair, in footnote 24 BCS agree with the Blatt, Butler, Schafroth group\cite{BBS2,Sha} in pointing out that the quantum coherence of the pairs over the entire sample is crucial to the superfluid properties. Nonetheless, the net effect of the arguments of both BCS and Landau is to reject the perfectly reasonable position that understanding the condensate in the Bose-Einstein situation will illuminate the behavior of the superflows in both helium and also superconductors.       

The obvious question is whether in downplaying the effect of the condensate wave function,  BCS leave anything out.  In one sense they do not.  They do point out that pairing of electrons with total momentum equal to ${\bf q}$ rather than zero will lead to current carrying states. (See \se{superflow} below.) So they have explained, almost in passing, the existence of supercurrents.  However, later on, this explanation seems to have disappeared.  On page 177 they say
``There is thus a one to one correspondence between excited states of the normal  and the superconducting phases.''
This statement does not work for the states carrying supercurrents around a ring.  These can also be viewed as excited states of the superconducting phase, and they have no direct analog in the normal phase of an electron gas.   

Further BCS do not seem to carry their discussion of the current-carrying states very far.  Specifically they do not point out that  in the presence of the current, the gap function, $\Delta$, becomes complex and can vary in space as $\exp(i{\bf q}\cdot\bf{r}/\hbar)$.   The reader would be helpfully guided by such a statement.  It might even have been helpful to BCS themselves.   It seems very strange that neither the contemporary audience nor BCS  asked for strong and convincing arguments for the amazing observed stability of the flow of supercurrents. 

The omission of the metastable current carrying states makes the gauge invariance of the BCS paper  problematical.  Gauge invariance, a basic symmetry of quantum physics, involves the fact that changing the phase angle of quantum wave functions as 
$ \Psi(\r) \rightarrow  \exp(i\a(\r)) \Psi(\r) 
$
leaves the quantum system unchanged if one also changes the electromagnetic vector potential, $A(\bf{r})$, in a suitable manner.  The vector potential, $A(\bf{r}),$ appears in BCS; the phase angle of the wave function, $\Delta$, does not. This omission leaves the question of whether BCS obeys this basic invariance of quantum physics in limbo.   At the time, theorists argued that BCS must be wrong because of its lack of gauge invariance.   

However, soon after BCS, in a very important paper, P.W. Anderson showed how the gauge invariance works\cite{Anderson}.   Oscillations of the gap parameter, or equivalently of the condensate wave function, produces extra states of the system, states which rescue the gauge invariance of the BCS theory.  This new type of excitation is the theoretical grandfather of the Higgs boson of particle physics.

\section{Condensate Studies}
\subsection{Ginzburg-Landau: 1950}\la{Ginzburg-Landau}
Well before BCS, two Soviet authors, Landau and Vitale Ginzburg\cite{GL}, had an excellent insight into the possible macroscopic  behavior of superconductors.  They applied the Landau theory of phase transitions\cite{Landau1935, Landau1937} to superconductors.  Two conditions are required for this application.  One condition was that the ``order parameter''  that described the amount of ordering produced by the superconducting phase transition had to be relatively small. The other was that mean field theory was applicable. In superconductors, the latter condition will ensue when the bound pairs are large enough so that many electrons interact at once.    This overlapping pair criterion was very well satisfied in the kind of superconductor known in and before the 1950s. The small order parameter criterion was well satisfied for temperatures near the critical temperature for this phase transition.\footnote{In unpublished work, Kurt Gottfried and I showed that the region of applicability of mean field theory does not include a small range of temperatures near the critical temperature for onset of superconductivity. However, the range of non-applicability is so narrow as to be irrelevant in almost all studies of simple superconducting metals. }  The only other requirement for the application of the Landau phase transition theory was to get the right symmetry for the order parameter.

These authors assume and assert that the order parameter has the symmetry and behavior of a quantum wave function.  Accordingly they take the order parameter to be a complex number, $\Psi$.   They borrow a piece of the  Landau theory of second order phase transitions which would give for temperatures, $T$, near the critical temperature, $T_c$ an equation for the order parameter of the form: 
\be
[~a (T-T_c)  + b |\Psi|^2~]   \Psi=0 
\la{ptTheory}
\ee       
with $a$ and $b$ being undetermined constants. The appearance of the $ |\Psi|^2$ is appropriate for a situation in which the order parameter, $\Psi$, is a complex number and has a physically undetermined phase angle.  Note that the form of  \eq{ptTheory} like the form of the 1937 Landau phase transition theory\cite{Landau1935, Landau1937} has no accommodation for any spatial dependence of  the order parameter.  It would, I believe, be natural to make provision for a spatial dependence of the order  by including a term in $\nabla^2 \Psi$ in \eq{ptTheory}\footnote{In my papers\cite{LPK09,LPK11}, I incorrectly assumed that the 1935-7  Landau work did make provision for the spatial dependence of the order parameter by including a term in $\nabla^2 $ applied to that parameter.  It did not.  }.   This extra term would bring this Landau work into line with the classical work of Ornstein and Zernike\cite{OZ} on the liquid gas phase transition.  However, in their 1950 work Ginzburg and Landau do include the possibility of spatial dependence.   

They do this by  combining \eq{ptTheory} with the Schr\"{o}dinger equation for the quantum theory of a particle with energy, $E$,  in the presence of a electromagnetic vector potential, $A({\bf  r} )$,
\be
\frac{ [-i \hbar \nabla-qA({\bf  r} )/c ]^2}{2m}
   \Psi({\bf r})=E \Psi({\bf r}) 
   \la{Schrodinger}  
   \ee 
Here $c$ is the speed of light, $\hbar$ is Planck's constant while $q$ and $m$ are the charge and mass of the particle. When combined the equations become the result now known as the Ginzburg-Landau equation:
\bsubs\la{GL}
\be
\{~a (T-T_c)  + b |\Psi(\r)|^2~+ [ -i \hbar \nabla-qA({\bf  r} )/c ~]^2/(2m)] \}  \Psi( \r)=0
\la{GL1}
\ee
We need also an equation for the electromagnetic current. which then takes the form 
\be
J=-i \hbar ( \Psi^* \nabla \Psi -\Psi \nabla \Psi^*)  -\frac{2q}{c}  |\Psi(\r)|^2 A(\r)
\la{GLJ}
\ee
\esubs 

Since the work is purely phenomenological, $a,b,q,$ and $m$ are all unknown, but the form of the equation looked promising.

Ginzburg and Landau also described a microscopic definition of the condensate wave function in terms of the quantum mechanical density matrix.  This definition  agrees with the one  introduced in parallel by Oliver Penrose\cite{ODLRO}, which then, came to be named off-diagonal-long-range-order (ODLRO).  ODLRO is considered today to be a defining characteristic of superfluids. The paragraph\cite{GL}[page 550] in which this definition is given is quite remarkable.  It is marred by a crucial misprint.\footnote{This paragraph and this misprint was pointed out to me by Pierre Hohenberg.}  In contrast with the usual confidant language of the authors, it is filled with tentative words: ``consider'', ``suppose'', ``it might be thought'', ``it is reasonable to suppose''.   It is as if the authors could not agree on simple declarative statements.   Perhaps this tone is part of the reason why BCS did not pick up on the Ginzburg-Landau paper as a definition of important characteristics of a superconductor.  Further the tone partially explains why Ginzburg-Landau is not generally cited as a source of ODLRO.

In 1962,  Lev Gor'kov\cite{Gor'kov2} was able to derive the Ginzburg-Landau equation from the BCS theory, obtaining a $\Psi$ proportional to $\Delta$.  This derivation then gives the values of the various constants in \eq{GL}.  Specifically, he found that the charge in question, $q$, is twice the electron charge.  This is the value we might have expected since the wave function describes bound pairs of electrons.  At this point, we finally have a conceptually complete theory of the kind of superconductivity that was known in the 1960s and before.\footnote{Specifically, BCS had constructed a theory of weak-coupling electronic superconductivity in situations with time-reversal invariance. This theory must be modified for other cases.  For example, it fails in the presence of magnetic impurities and also for the subsequently discovered high temperature superconductors.  Nobody knows how to do the analog of BCS for these  high $T_c$ materials.}  However, a complete theory is not yet a complete understanding.  It is not clear that in 1962 anyone yet had a good intuitive understanding of the implications of the condensate wave function.       

\subsection{Abrikosov: 1952, 1957} 
The first use of the new work was A. A. Abrikosov's application of the Ginzburg-Landau equation to the behavior of superconductors in a magnetic field\cite{Abrikosov1, Abrikosov2}.   The first of these papers described the behavior of films and introduced the idea that there were two kinds of superconductors, Type I and Type II, with the latter being a new kind introduced by Abrikosov. This kind of superconductor has novel properties because it tends to break up into  normal and superconducting regions.   The second of the papers  showed that in a magnetic field type II materials formed vortices, swirls of supercurrent surrounding normal regions. (See \fig{TypeII}.)  Landau termed this behavior ``exotic'' and at his suggestion this work remained unpublished for several years\cite{AbrikosovPT}.  After  Feynman had published his vortex work\cite{Feynman}, Landau relented, Abrikosov finally published\cite{Abrikosov2}, and the work became a major contribution to superconductivity theory.\footnote{Feynman's vortex predictions\cite{Feynman} were preceded by Lars Onsager's\cite{Onsager} and followed by estimates of the size of vortex effects in helium by Penrose and Onsager\cite{PO}.}  Once again,  it would appear, Landau undervalued work involving the condensate wave function.

\begin{figure}
\begin{multicols}{2}
\includegraphics[height=4.5 cm ]{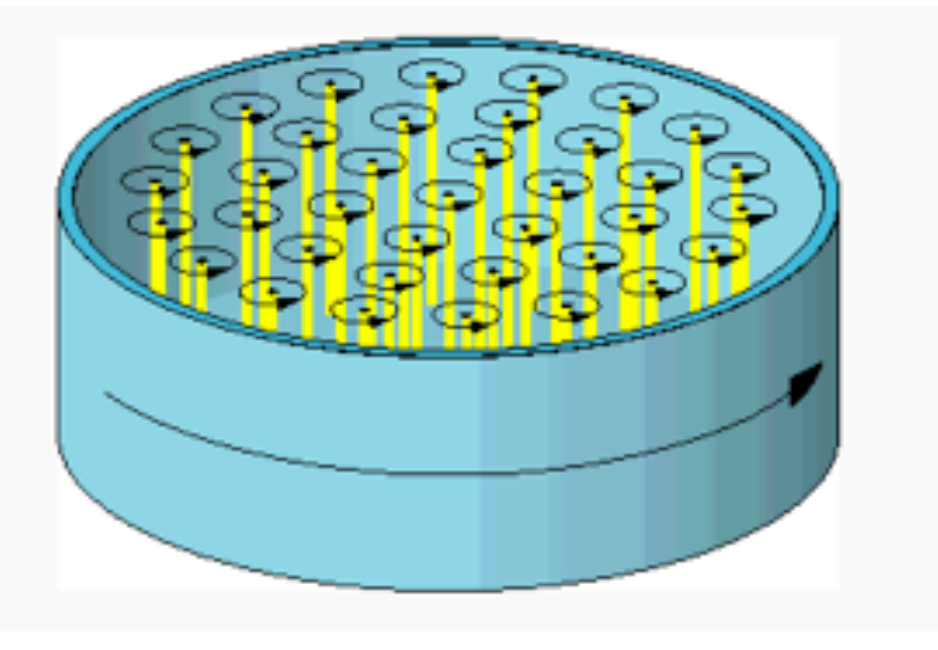}
\caption{Type II superconductor.  A magnetic field produces vortices each containing a quantized unit of magnetic flux.     }
\la{TypeII}
\end{multicols}
\end{figure} 

In this context it is interesting to note that the 1957 BCS article does not quote either the Bogoliubov work of 1947\cite{Bogoliubov},  the Ginzburg-Landau paper from 1950\cite{GL} or the 1952 Abrikosov work\cite{Abrikosov1}.  In fact, no Soviet article is quoted.  This is perhaps an indication of the bad Soviet-American relations in that period, or more directly of the bad circulation of Soviet work in the U.S..  Or perhaps it is one more indication of the BCS downplaying of the importance of condensate effects. 

But the most important condensate effect is yet to come.

\subsection{Josephson: 1962}
In 1960, Ivar Giaevar found that he could see the structure of the quasiparticle spectrum of superconductors. He used an electrical voltage to pull electrons from the superconductor, through a narrow gap of insulating materials, and thence into another metal\cite{Giaevar}.  The quasiparticle spectrum could be inferred from a measurement of the current as a function of the voltage across the gap.   The first theories of this effect\cite{B2,CFP} left out the lowest order effect of having the condensate wave function tunnel across a barrier separating two superconducting materials.  A further paper on the subject by Brian Josephson\cite{Josephson} argued that   the tunneling of the condensate wave function  would produce an additional measurable current, and predicted the size of the current.   The specific predictions were very surprising:
\begin{itemize}
\item Time Independent Effect: In a situation in which the voltage is the same on the both sides of the barrier, there is nonetheless a possibility for a supercurrent to tunnel across the barrier. The supercurrent is driven by difference in the phase angles, $\phi$, between the wave functions, $\Psi$ on the different sides of the gap.  The current then has the value $J=J_0 \sin( \delta \phi) $, where $ \delta \phi $  is that difference in phase angle.  In this way, the wave functions on the two sides of the gap are made directly visible.
\item Time Dependent Effect:  Correspondingly when there is a voltage, $V$, across the gap, there is a time dependent current with a value
$J=J_0 \sin( 2 e V t /\hbar) $.  This dramatic effect offers the possibility of a measurement of the charge on the electron, $e$, via a simple measurement of a voltage and a frequency.
\end{itemize}
\begin{figure}
\begin{multicols}{2}
\includegraphics[height=4.5 cm ]{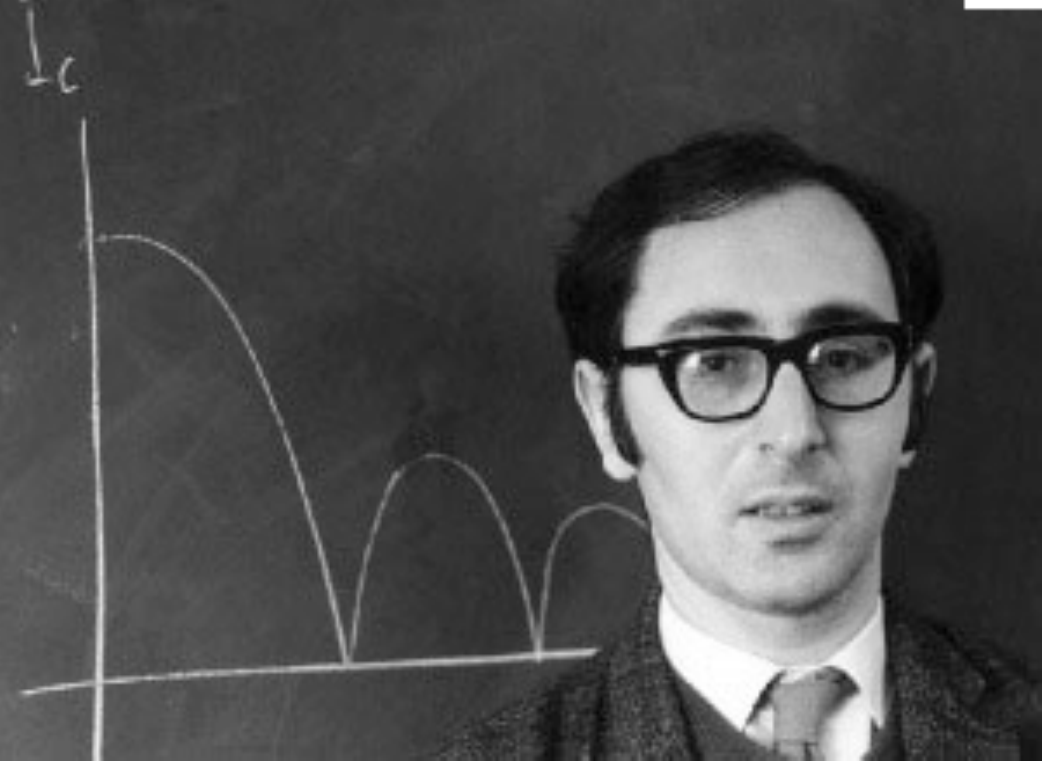}
\caption{Brian Josephson was a graduate student when he described the result of wave-function tunneling  between two superconductors.  In a {\em Josephson junction}, the pair wave function  tunnels from one side of the barrier through to the other side and interferes with the wave function on the other side. This interference produces a current.  }
\la{Josephson}
\end{multicols}
\end{figure} 

Bardeen did not believe that Josephson's calculations were correct\cite{Bardeen1962}, most probably because he did not deeply understand the implications of the existence of the condensate wave function and of the Ginzburg-Landau theory. 

In a famous debate at the Eighth International Low Temperature Physics Conference at Queen Mary College London  in September 1962, Josephson described his work while Bardeen argued that Josephson was wrong.    As the experimental facts came in, it became apparent that Josephson's predictions were, in fact, correct.     At the next low temperature conference, Bardeen acknowledged that Josephson had indeed been right.  

 In a series of papers written just before\cite{B0} and just after BCS\cite{B1,B2,B3,Bardeen1962}, one can see Bardeen working to absorb and use the new concepts\cite{B1,B2,B3}. Part of this is an attempt to define a useful tunneling theory\cite{B2,Bardeen1962}. Part is an attempt to use concepts related to condensate wave functions\cite{B0,B1, B3}.  By the time Bardeen recognized flux quantization\cite{B3}, he also recognized the existence of condensate wave functions.  He even followed Ginzburg-Landau and wrote down\cite{B3} a condensate wave function himself.  However, his disbelief\cite{Bardeen1962} of Josephson's work is sufficient evidence that, in 1962, he had not yet absorbed all the implications of the physical argument that stretched forward from Einstein's theory.

\section{Understanding superflow}\la{superflow}
\subsection{Understanding gained}\la{gained}
Like Josephson, we have all the ingredients in front of us so we can describe why superfluids and superconductors exhibit flow without frictional slowing.   This understanding arises from a consideration of the superfluid wave function, $\Psi(\r)$.

\subsubsection{Phase angles}
 Let us imagine for a moment that there were some wave function $\Psi(\r)$ describing a finite fraction of all the particles in a superfluid. Such a wave function would, of course, be complex and have the form
\be
 \Psi(\r)= \sqrt{\rho(\r)} \exp (i \a(\r))
 \ee 
Here  $\rho(\r)$ can be interpreted as being proportional to the probability of finding a particle at $\r$ or of the density of particles at $\r$ depending upon whether we interpret this wave function as describing one particle or the entire condensate.  In a non-relativistic situation, the phase angle, $\a(\r)$, is related to the velocity and the momentum of the particle(s) by 
\be
{\bf p}(\r)=m{\bf v}(\r)= \hbar \nabla \a(\r)
\la{velocity}
\ee
We can then give a physical interpretation to the variation of phase angle from place to place, but we cannot give a physical interpretation to the actual value of $\a(\r) $.  In the presence of a vector potential, $m{\bf v}(\r)$ is replaced by $m{\bf v}(\r)-q{\bf A}(\r) |\Psi(\r)|^2/c$.  The velocity is then gauge invariant while the momentum, ${\bf p}$, the vector potential $A $, and the phase angle, $\a$, are not gauge invariant.  To explain the Meissner effect, one has to describe the behavior of the vector potential and get into electromagnetic theory.   Instead of doing this, I shall concentrate on the behavior of superfluid helium. 

In helium these is no magnetism and no vector potential.  We identify the velocity of \eq{velocity} as the superfluid velocity and note that is it automatically the gradient of a scalar, the scalar being phase angle.  Landau\cite{L1}[page 314, where Landau refers to the ``Euler paradox''] has pointed us to fluid mechanics and D'Alembert's ``paradox'' which says that such a fluid velocity distribution cannot produce a net drag force on a body\cite{D'Alembert,GPF}.  So the condensate approach automatically gives frictionless flow.  However in not admitting a condensate wave function, both BCS and Landau have avoided \eq{velocity} and its direct connection with superfluidity.  

\subsubsection{Toroidal geometry}
Let us consider a thin tube of helium twisted into a torus as in \fig{RaceTrack}.   We can describe this situation with a wave function with a single coordinate, $\phi$, describing angular position in the tube.  Since the wave functions must be single valued the possible static wave functions are of the form
\be
\Psi(\phi)  = \sqrt{N_0}~ \exp(iM\phi). 
\ee
Here $M$ is a quantum number describing the angular momentum of the flow. Because the wave function must be single-valued, $M$ must be an integer.  If $M=0$, there is no flow; any other value gives a quantized flow.    Note that this description satisfies the Bloch's theorem requirement  (See \se{Background})   that the ground state ($M=0 $) carry no current.  

Because this flow encompasses a finite fraction of all the particles in the system it is very hard to stop.  Indeed quasiparticle scattering mechanisms like the ones envisioned in the Landau criterion can only produce a finite rate of scattering out of the condensate.  However, according Einstein's description of the mechanisms for boson scattering\cite{Einstein1916},  this very same scattering mechanism will return, on the average, an equal number of particles per unit time into the condensed state.  Because the condensate contains a huge number of particles, the fluctuations engendered by this process remain small compared to the occupation of the condensate so that these scatterings do not threaten the constancy of the values of $M$ or of the superflow.

If the circular channel holding the helium is relatively broad, the superflow can continue almost indefinitely.  But not quite.  An Abrikosov-Feynman vortex can move through the channel from inside to out or vice versa.  This vortex will carry one unit of angular momentum and change the value of $M$ by $\pm 1 $.  If one waits long enough, such tunneling of vortices will make the superflow decay to zero\cite{LF,LA}.  But, this tunneling is essentially a macroscopic process and can be quite slow.  In realistic cases, the superflow may well be stable for periods longer than the age of the universe.  

\subsection{Understanding Postponed} \la{NotInverntedHere}   
We can certainly conclude that both Landau and BCS are playing down the importance of the condensate dynamics in comparison to the quasiparticles dynamics.  We should ask why they engaged in this apparently counterproductive behavior. To some extent this outcome may be manifestations of the ``not invented here'' syndrome in which each scientist and group emphasizes what it has created and deemphasizes the creations of others.  There may well have been an unwillingness to admit something as radically different as a wave function extending through the whole body of the material.  Further, these world-leading authors may have been unwilling to admit that relatively uncelebrated scientists may have led the production of a quite radical solution to these very important physics problems. 

A more historical argument is suggested in a thesis by Edward Jurkowitz\cite{Jurk} who looked back at the development of quantum theory starting with the old guard: Einstein, Schr\"odinger, Weyl and moving on to new voices: Heisenberg, Pauli, Dirac, Jordan.  The old guard spent considerable effort in constructing unified field theories which often contained an allover wave function describing space-time.  They were unable to make these theories work. The newer group preferred to work with operator commutation relations. They were then very suspicious of any wave function which pretended to describe an entire physical system.  Einstein's condensate wave function did just that.   So perhaps people who had contact with the new voices in the history of quantum theory, like Landau and Bardeen, felt similar suspicions. 

One can easily imagine midcentury authors trying to make sense of their research by enforcing an almost complete separation between the microscopic and the   macroscopic.  They might be willing to admit of the familiar connections given by the microscopic determination of thermodynamic functions and transport processes, but being unwilling  to admit of new and exotic connections, like spatially varying order parameters or wave function extended through the whole body of the material. 

A fourth reason is recounted by Alan Griffin\cite{AG}. After the Einstein paper\cite{Einstein}, George Uhlenbeck\cite{UhlenbeckT} wrote a thesis in which he said that the Bose-Einstein transition was nonsense because it could only occur in an infinite system.    Uhlenbeck retracted his criticism\cite{Uhlenbeck} in 1938.  Nonetheless Uhlenbeck's objection perhaps turned people away from the Einstein work.  

I believe that the noise and bombast of the critical comments in Landau and BCS's work indicated  that at some level they were aware that they were making a mistake.  It has long been considered suspicious when people ``doth protest too much.''

\section{Fritz London:  1900-1954  }
In all the years following his first work on superconductivity in 1935, Fritz London was pushing in the right direction.  Working with his brother Heinz London, the outcome was an early and correct description of the electrodynamics of superconductors\cite{FHLondon,KG}.   Fritz London recognized the close analogy between superconductors and superfluids as well as the connection between these problems and Bose-Einstein condensation.  His accomplishments can be found in his  books on superconductors and superfluids\cite{London1,London2} 

By no means did London get the sort of recognition that was tendered to Landau, BCS or the other leading lights.  London was a Jew and a German in difficult times.  Fleeing Germany, he could not get a long-term job in England and had to work hard to finally find a job at Duke University\cite{KG}.  

But recognition did finally come. In 1953, the Royal Netherlands Academy of Sciences gave him the highest recognition of Netherlands Science, its Lorentz medal.  Previous winners were Planck, Debye, Sommerfeld, and Kramers.     In the same year, his University recognized him with the title of ``James Duke Professor''. 

The most significant recognition, however, came from John Bardeen.  In 1972, John contributed monies to be used for the Triennial Award in Low Temperature Physics, which would thenceforth be the Fritz London Award.  He also set up a lectureship in London's name at Duke University.   In 1990, John wrote a piece, an afterward,  for Kostas Gavroglu's {\em Fritz London, a scientific biography}\cite{KG}[pp. 267-272], saying, among other things, 
\begin{quotation}
By far the most important step in understanding the phenomena [of superconductivity] was the recognition by Fritz London that both superconductors and superfluid helium are macroscopic quantum systems   [...]  The key to understanding superfluidity is macroscopic occupation of a quantum state. 
\end{quotation}
\subsection{Afterward}
However, despite this recognition for Fritz London no similar recognition has been forthcoming for the objects of Landau's and BCS's comments: Tisza and Blatt, Butler, Schafroth.   

Physicists became comfortable in working with coherent states of many bosons through experimental maser and laser studies of the late 1950s and the coherent state analyses of Julian Schwinger\cite{Schwinger} and Roy Glauber\cite{Glauber}.  The Bose-Einstein condensation, abbreviated BEC, has been directly studied in the behavior of cold atomic gases.   This condensation is accepted as the basis of all superfluid behavior.

\section*{Acknowledgments}
This work was partially supported by the University of Chicago NSF-MRSEC under grant number DMR-0820054.   I have had instructive conversations on the topics of this paper with David Pines, Silvan Schweber, Gloria Lubkin, Gordon Baym, Edward Jurkowitz, Margaret Morrison, Joel Lebowitz, Roy Glauber, S\'{e}bastien Balibar, and Paul Martin.
\newpage{}

\end{document}